\documentclass[12pt]{iopart}
\usepackage{graphicx}
\usepackage{bm}
\usepackage[hang,loose]{subfigure}
\usepackage{wasysym}
\begin{document}
\title[High resolution UCNP ionization]{High Resolution Ionization of Ultracold Neutral Plasmas}
\author{P. McQuillen, J. Castro,
T. C. Killian}
\address{Department of Physics and Astronomy and Rice Quantum Institute, Rice University,
Houston, TX 77005, USA}
\ead{patrickmcquillen@rice.edu}
\begin{abstract}
Collective effects, such as waves and instabilities, are integral to our understanding of most plasma phenomena. We have been able to study these in  ultracold neutral plasmas by  shaping the initial density distribution through spatial modulation of the ionizing laser intensity. We describe a relay imaging system for the photoionization beam that allows us to create higher resolution features and its application to extend the observation of ion acoustic waves to shorter wavelengths. We also describe the formation of sculpted density profiles to create fast expansion of plasma into vacuum and streaming plasmas.
\end{abstract}

\maketitle

\section{Introduction}
Collective modes are central to our understanding of plasma dynamics and transport properties \cite{sti92,ich04}. They often lead to instabilities \cite{mel86} that can dominate plasma dynamics and create significant complications for practical applications such as magnetic confinement fusion.
In principle, and for most regimes of plasma density and temperature,
these phenomena are well described within classic theoretical
frameworks such as  continuum fluid or kinetic descriptions, but the possible complexity in plasmas is continually leading to the observation of interesting new phenomena. In addition, at relatively high densities and/or low temperatures, when plasmas become
strongly coupled, or non-ideal \cite{ich82}, plasma dynamics becomes
hard to describe theoretically or simulate numerically, so new experiments  probing this regime are of great fundamental interest and important for understanding the behavior of matter under extreme conditions of energy and density \cite{mur04}.

Recent advances in plasma creation have allowed for the excitation of ion collective modes in ultracold neutral plasmas~(UCNPs) \cite{cmk10}. UCNPs \cite{kpp07,kil07} are formed by photoionizing laser cooled atoms near the ionization threshold, and they access an exotic regime of plasma physics in which electron and ion temperatures are orders of magnitude colder than in traditional neutral plasmas.
 These plasmas have unique properties arising from their ultracold temperature, small size, and expansion dynamics. They also display correlated particle dynamics \cite{mur01,scg04,csl04,cdd05,mur06PRL,ppr05PRL} that reflect strong coupling \cite{ich04} of the constituents, which allows them to probe a regime of nature in which correlated, many-body interactions are dominant.

In strongly coupled systems, the  Coulomb interaction energy exceeds the thermal energy, as characterized by a Coulomb coupling parameter, $\Gamma\geq1$, where
\begin{equation}
\Gamma=\frac{e^{2}}{4 \pi \varepsilon_0 a k_B T}.
\label{eqcoulomncouplingconstant}
\end{equation}
Here $T$ is the temperature and $a$ is the Wigner-Seitz radius,
$a=\left[3/(4\pi n)\right]^{1/3}$ for density $n$.  Because of the low temperature in UCNPs, ions equilibrate
with $2< \Gamma_i < 5$ \cite{scg04,cdd05}. It is possible to set initial
conditions for large electron $\Gamma_e$, but heating processes involving many
different collisional effects will lower the coupling to $\Gamma_e \leq 0.2$
\cite{rha03,kon02,mck02,gls07,fzr07} within a few microseconds. For $\Gamma>1$, particle correlation effects become important.

 A wide range of phenomena has been studied in UCNPs, such as expansion of the plasma into surrounding vacuum with \cite{zfr08expansion} and without \cite{lgs07,gls07,rha02} an applied magnetic field, adiabatic cooling of electrons during expansion \cite{zfr08expansion,fzr07}, three-body recombination at ultracold temperatures \cite{klk01,fzr07,pvs08}, spontaneous evolution of a frozen Rydberg gas into an ultracold plasma \cite{gpr03}, equilibration of ions after plasma creation \cite{csl04,ccb05}, and
disorder induced heating of electrons \cite{kon02,mck02} and ions \cite{csl04,ccb05,mur01,ppr05PRL} and kinetic energy oscillations \cite{csl04,ccb05,lcg06}
due to effects of strong coupling.
Langmuir oscillations  have been resonantly excited and used as a probe of plasma density \cite{kkb00,fzr06}, and
 an $\mathbf{E}\times \mathbf{B}$ electron drift instability was also observed \cite{zfr08}.

Recently, Castro \textit{et al.} \cite{cmk10} used a new technique involving spatially masking the photoionization-laser intensity to shape the initial density distribution of UCNPs and excite long-wavelength ion acoustic waves (IAWs). The excitation wavelength was as short as $\lambda\approx500\,\mu$m with the limit set by diffraction effects. Shorter wavelength excitations would access more interesting physics, such as the transition from a sound wave to a dispersionless ion plasma oscillation when the wavelength becomes on the order of the Debye screening length, $\lambda_D=(\varepsilon_0 k_B T_e/
n_{e} e^2)^{1/2}$, where $T_e$ and $n_e$ are the electron temperature and density respectively.
 Furthermore, corrections to the IAW dispersion relation due to strong coupling of ions are predicted for $\lambda \apprle a$
  \cite{wbh97,rka97,kse98,mur98,mur00PPlasma,kaw01,ber93} and sharp features in the ion density profile should also lead to large ion accelerations that may excite shock waves \cite{rha03,kpp07} and solitons  \cite{nbs99}. In this paper we present improvements to the ionization system that decrease the observable excitation wavelength to $\sim$100\,$\mu$m, approaching the Debye length under accessible UCNP conditions and  only 10 times the Wigner-Seitz radius.

\section{Experimental Setup}

Strontium UNCPs are created in a magneto-optical trap (MOT) containing $\sim300$M laser cooled $^{88}$Sr atoms by photoionizing approximately half of the atoms near the ionization threshold with a pulsed narrowband laser \cite{kpp07}. The density distribution of the atoms  is a spherical gaussian, $\propto \mathrm{exp}(-r^2/2\sigma^2)$, with characteristic radius $\sigma\sim 1-2$\,mm, and if the ionization beam intensity profile is uniform, the plasma has a similar form. The electrons have a tunable temperature, ($\frac{2}{3}E_e/k_B\sim1-100$\,K), determined by the energy difference ($E_e$) between the ionizing photon and the ionization potential. The ions, with their relatively large mass, inherit the temperature of the laser-cooled atoms ($\sim 10$\,mK).

\begin{figure}[ht]
\begin{center}
\includegraphics{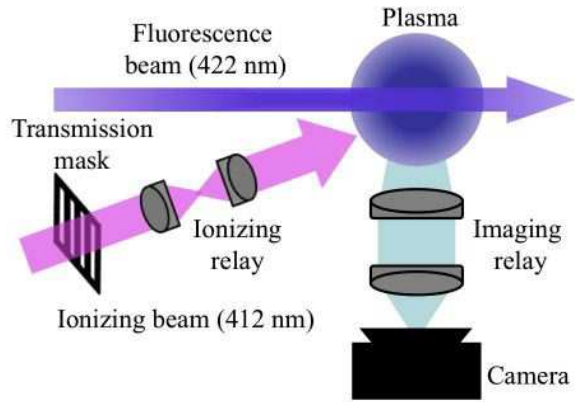}
\caption{\textbf {Experimental setup} showing the optical arrangement for ionizing and imaging an UCNP. The ionizing relay lenses are set at a distance for optimum imaging of the transmission mask onto the plasma, and the transmission masks are mounted in rotation and translation stages. To increase the ionization fraction the ionizing beam can be rerouted for a second pass through the chamber (not shown). Placing the mask in the beam path just before the second pass creates smaller amplitude density perturbations.}
\label{fig:setup}
\end{center}
\end{figure}

Two-photon ionization is realized with one photon from a pulse amplified beam of light from the cooling laser, driving the Sr $^{1}\textrm{S}_{0}$ - $^{1}\textrm{P}_{1}$ transition at $461$\,nm, and a second photon from a pulsed dye laser, at $412$\,nm, which promotes the electron to the continuum. We use a mask on the second ionizing beam in order to spatially modulate the laser intensity. This simple technique allows one to spatially manipulate the ionization fraction, in turn controlling the initial density distribution of the UCNP. For instance masking half the beam with a straight edge  produces a single Gaussian hemisphere of plasma.
Further flexibility is provided by the option of reflecting the ionizing beam through the chamber a second time to increase the ionization fraction, and by masking only the second pass of the $412$\,nm ionizing beam we can create smaller amplitude perturbations. For example, a 100\% contrast square wave mask was used in this fashion in previous studies to excite low amplitude IAWs with user-selected wavelength  \cite{cmk10}. Although highly versatile, this straightforward technique is limited in terms of achievable  feature  size due to laser diffraction over the distance from the mask to the plasma. To overcome this limitation we have implemented a relay lens system  to image the mask onto the plasma with much higher resolution.

Our design consists of a matched, positive focal length lens pair (Thorlabs AC508-200-A $f=200$\,mm achromatic doublets) in a 1:1 conjugate ratio. To achieve maximum performance, many factors, such as lens aberrations, alignment, and aperture must be balanced against experimental design limitations, but this configuration was predicted with Lambda Research's OSLO optical design software to produce a contrast transfer function (CTF) of  50\% for  a 50\,cyc/mm pattern.
The lenses are held in place with a custom lens tube adapter that can be securely attached to the vacuum chamber. This direct mounting method ensures correct and reproducible mask-lens-plasma alignment (coaxial with ionizing beam and distance $f$ from the plasma). Transmission masks are mounted a distance $f$ back in a $\theta_z$ rotation mount atop an xz translation stage (laser propagation $\vec{k}/|\vec{k}|=\hat{z}$). Rotational alignment ensures that ionized edges are aligned vertically with the plasma imaging system's line of sight, x alignment allows centering of a pattern transversely on the plasma, while micrometer z adjustment allows fine positioning of the mask image so that the atoms are centered within the depth of field.

\section{Improved Resolution}
We measured the contrast transfer function of the ionization relay system by imaging a 1951 USAF resolution target onto a CCD camera. The CTF is calculated from the normalized signal (S) as $CTF(\nu)=(S_{max}(\nu)-S_{min}(\nu))/(S_{max}(\nu)+S_{min}(\nu))$, where $\nu$ is the spatial frequency of the square wave pattern. The system exceeded design goals, with a CTF of $\sim 100$\% at 20\,cyc/mm over the typical depth of field of the plasma $\pm2$\,mm and maximum focal plane resolution of CTF $\sim 100$\% at $>$80\,cyc/mm. Note, this later measurement was limited by CCD pixel size.

\begin{center}
\begin{figure}[ht]
\subfigure{
\includegraphics{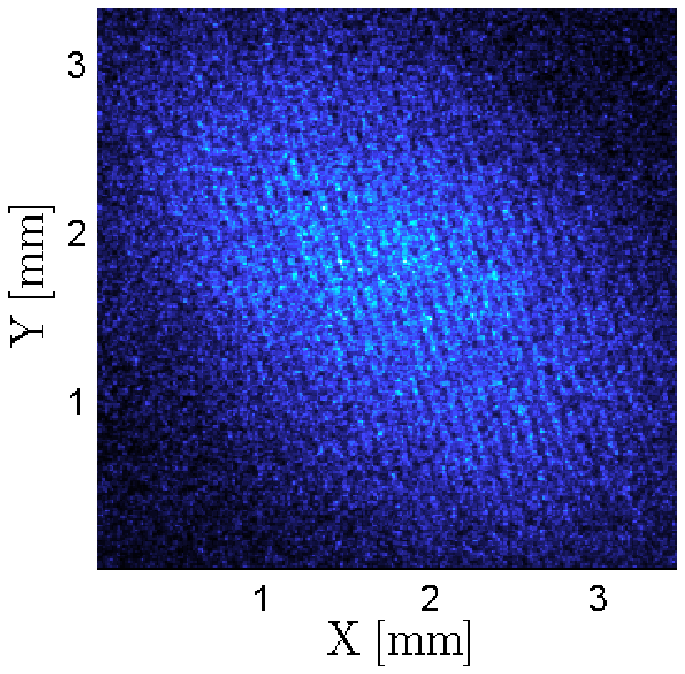}
\label{fig:single}}
\subfigure{
\includegraphics{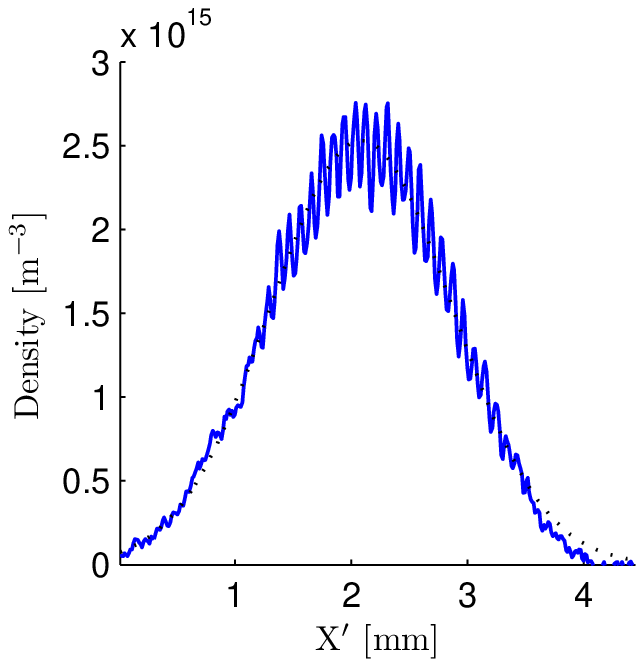}
\label{fig:guass}}
\label{fig:imaging}
\caption{\textbf{Fluorescence images and density profiles.}
(left)
On resonance fluorescence image of UCNP with 10\,cyc/mm modulation, 100\,ns after plasma creation. Limitations to optical access of the ionizing beam into the vacuum chamber produced the resulting oblique orientation of the plasma.
(right) Density profile of plasma (solid) with Guassian fit (dotted). The lack of contrast on the left side  results from reflections of the ionizing beam inside the chamber. $X'$ is the distance along a perpendicular to the density features.}
\end{figure}
\end{center}

In situ diagnostics of the ionizing relay system was accomplished with fluorescence imaging of the Sr$^{+}$ ions, on the principal transition, ${}^{2}\textrm{S}_{1/2}-{}^{2}\textrm{P}_{1/2}$ at 422\,nm \cite{cgk08} (Fig. \ref{fig:setup}). A sheet of light, with a beam waist of $625\,\mu$m, propagating perpendicular to the imaging system line of sight and $106^{\circ}$ away from the 412\,nm ionizing beam, is used to excite fluorescence. Relative plasma density in the illuminated region is measured directly by scanning the imaging laser over the doppler broadened resonance of the entire plasma cloud and summing the captured fluorescence in post analysis. Absolute calibration of density is accomplished with complimentary absorption imaging \cite{scg04}. From the 2D images we create 1D slices by cropping a narrow strip perpendicular to the density features and integrating along the narrow dimension (Figs. \ref{fig:single} and \ref{fig:guass}). For these measurements, a single pass of the ionizing beam through the chamber is used, which should lead to 100\% density contrast for well-resolved patterns.
 Using these analysis tools and square wave masks of various wavelengths we measured contrast of the imaged plasma density, which represents a CTF of the entire ionizing/imaging system.
 Figure \ref{fig:guass}, shows one such measurement where we achieve our highest resolution of CTF=13.5\% at 10\,cyc/mm.

 This resolution limit is significantly worse than what was measured for the photoionization relay during bench testing. Effects of plasma dynamics and misalignment of the planes of constant density with the line of sight of the imaging system cannot explain the discrepancy. We suspect that the imaging system is currently limiting our resolution and the plasma density features have higher contrast.  Improvements to the imaging system are now underway.

\section{Experiments with Sculpted UCNPs}


\begin{center}
\begin{figure}[ht]
\subfigure{
\includegraphics{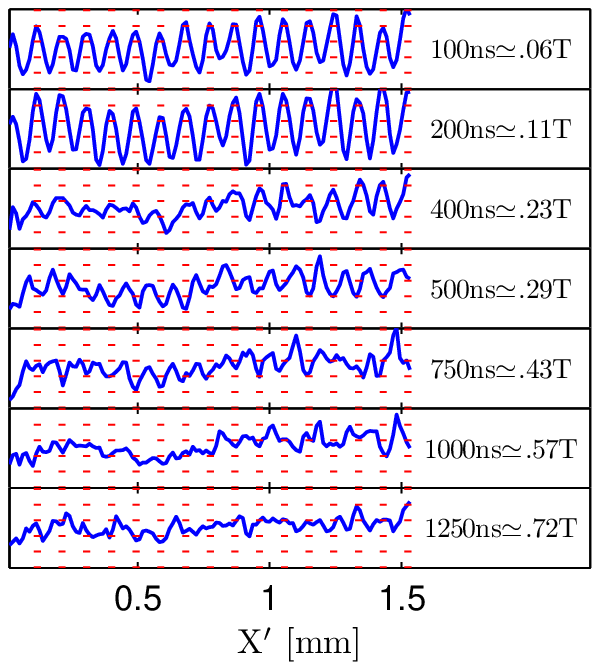}
\label{fig:evolution}}
\subfigure{
\includegraphics{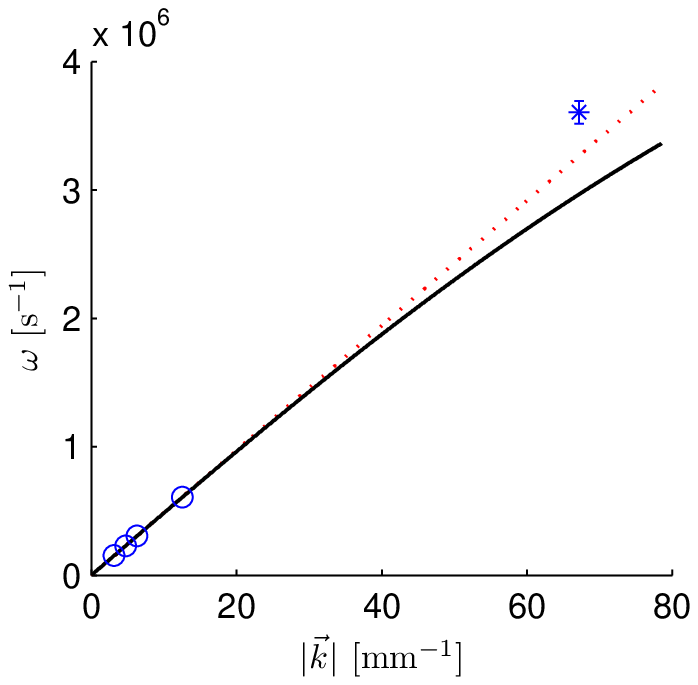}
\label{fig:omegaoverk}}
\label{fig:evol}
\caption{\textbf{IAW Evolution for 10\,cyc/mm perturbation and $T_e=25$\,K.}
(left) Evolution of the normalized perturbation. Note the initial periodicity closely matches that of the 10\,cyc/mm mask, with dashed lines indicating positions of amplitude peaks. Between 400 and 500\,ns the nodes and anti-nodes invert, which marks the first quarter cycle for the standing wave. A more rigorous fit, as described in the text, yields a quarter period for the IAW of $T/4=440 \pm 40$\,ns.
(right)
Extended IAW dispersion plot. The solid line corresponds to the theory curve, with linear extension of large wavelength regime shown  as a dotted line to emphasize the screening correction. Circles indicate previously measured data \cite{cmk10}, while the asterisk shows the new measurement.
}
\end{figure}
\end{center}

The observed resolution represents a five-fold improvement over the results presented in \cite{cmk10}, and we can now study plasmas with density modulations with spatial frequency as high as 10\,cyc/mm.  To isolate the density modulations, density profiles such as the one shown in Fig. \ref{fig:guass} are fit to a Gaussian profile and then divided by the fitted Gaussian.
With time resolved imaging, we observed the evolution of the ion density modulation (Fig. \ref{fig:evolution}),
which can be described in terms of IAWs similar to \cite{cmk10}. IAWs are longitudinal density modulations with slow enough ion dynamics for electrons to remain isothermal.  In an infinite homogeneous medium, IAWs have the following dispersion relation for frequency $\omega$ and wave vector $k=2\pi/\lambda$,

\begin{equation}
\left(\frac{\omega}{k}\right)^2=\frac{k_BT_e/M}{1+k^2\lambda_D^2},
\label{eq:dispersion}
\end{equation}
where $M$ is the ion mass.

For the short-wavelength IAWs that are now accessible, the higher frequency allows a simpler analysis than used previously \cite{cmk10}. The oscillation period is small compared to the plasma expansion timescale, therefore the electron temperature and IAW wavelength, which affect the oscillation frequency, do not change significantly during the wave period. This permits a simple description of the density perturbation as a constant wavelength, constant frequency, standing IAW, i.e $\delta n=A_0e^{-\gamma t}\mathrm{cos}(kx)\mathrm{cos}(\omega t)$, where $A_0$ is initial amplitude, and $\gamma$ is an exponential decay rate. A fit of this model to data shown in Fig.\ \ref{fig:evolution} yields a radial oscillation frequency of $\omega=3.6(1)\times 10^{6}\,\mathrm{s}^{-1}$ and a decay rate $\gamma=9(1) \times 10^{5}\,\mathrm{s}^{-1}$.

Figure \ref{fig:omegaoverk} shows the comparison of measurements and theory for the dispersion relation. At higher resolution, the wavevector is in the transition regime from linear, acoustic behavior to the limiting case of a dispersionless ion plasma oscillation with $\omega_{pi}=\sqrt{n e^2/M \varepsilon_0}$ for large $k$. With higher electron temperature, which is easily accessible, we can push further into this regime. The measurement, however,  does not show the expected rollover. This could be due to the initial density modulation being much larger than in previous studies \cite{cmk10} because equation \ref{eq:dispersion} is only valid when the excitation is a small perturbation to the density distribution.

The observation of fast damping of the oscillation is also an interesting observation. The damping is much stronger than observed at longer wavelengths \cite{cmk10}, which may indicate the influence of Landau damping since this effect becomes large as the wavelength decreases \cite{bth10}. Variation of frequency across the sample due to variation of density and $\lambda_D$ may also contribute to the decreasing amplitude. With our increased resolution, we can now study damping in detail.

\begin{center}
\begin{figure}[ht]
\subfigure{
\includegraphics{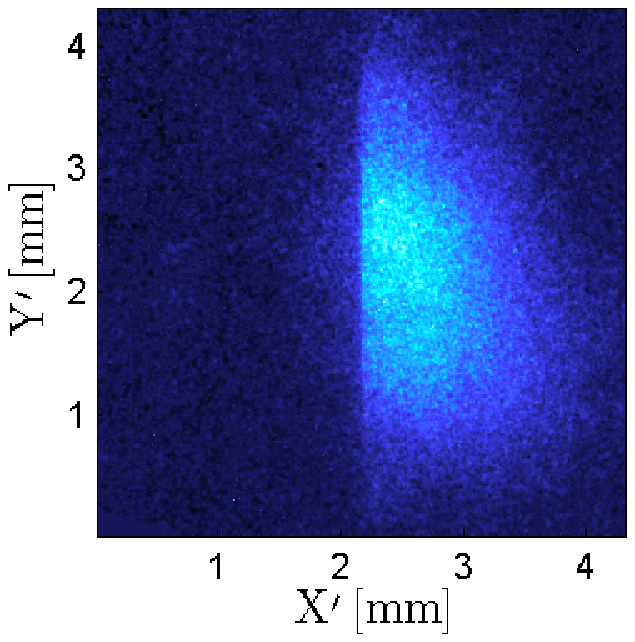}
\label{fig:hemisphere2d}}
\subfigure{
\includegraphics{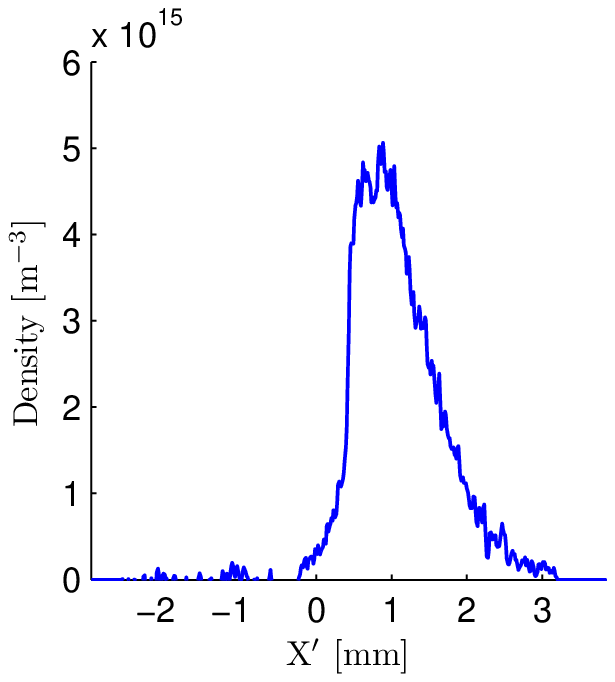}
\label{fig:hemisphere}}
\caption{\textbf{Single hemisphere}
(left) 2D image of a plasma hemisphere. Half of the ionizing beam is blocked to produce one hemisphere of an otherwise spherical Gaussian plasma.
(right)
1D density profile of a plasma hemisphere. Preliminary studies show how the strong gradient produced here results in ions that are accelerated strongly into the vacuum.
}
\label{fig:hemi}
\end{figure}
\end{center}

Beyond IAW dispersion and damping, we can also study a variety of situations in which plasmas are strongly accelerated into the vacuum or into another plasma. Large plasma density gradients, accessible with high resolution photoionization, produce large plasma accelerations and velocities, enabling the study of nonlinear effects such as shock formation and solitons \cite{mel86,rha03,kpp07,nbs99}. The importance of the density gradient can be seen from a hydrodynamic description using the coupled ion-electron momentum balance equations at early times when ion velocities are small \cite{kkb00}. Acceleration of the ions ($a_0$) is provided by the electron pressure ($P_e$) according to $Mn\vec{a_0}\approx-\vec{\nabla} P_e\approx-\vec{\nabla}\left[nk_BT_e(0)\right]$. $T_e(0)$ is expected to be close to uniform \cite{rha03}, leading to acceleration that varies as $\vec{\nabla} n$.
For an unperturbed UCNP, the density profile is spherical gaussian with $|-\nabla n |\approx n/\sigma_0 $, giving the characteristic radial acceleration on a timescale of about $10\,\mu$s that has been studied extensively in the context of self-similar plasma expansion \cite{kpp07}.
 However, when sharp features with length scale $l\ll\sigma_0$ are imprinted on the plasma density, the local gradient and resulting plasma acceleration and velocity can be much larger.

\begin{center}
\begin{figure}[ht]
\subfigure{
\includegraphics{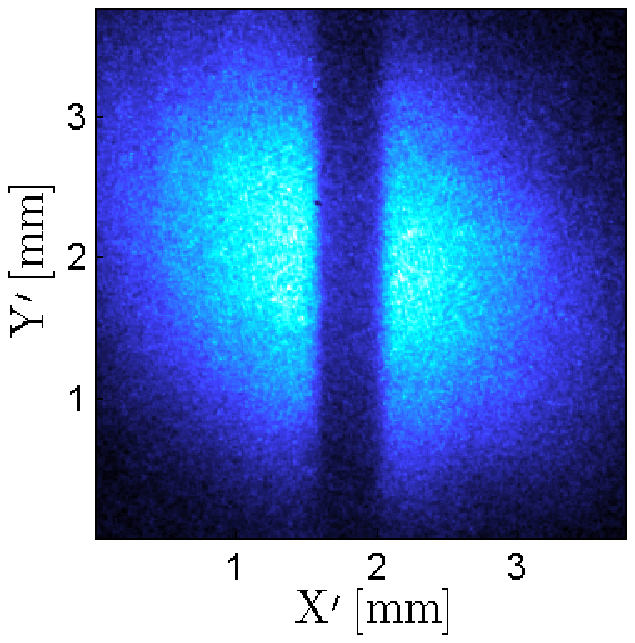}
\label{fig:streamplasma2d}}
\subfigure{
\includegraphics{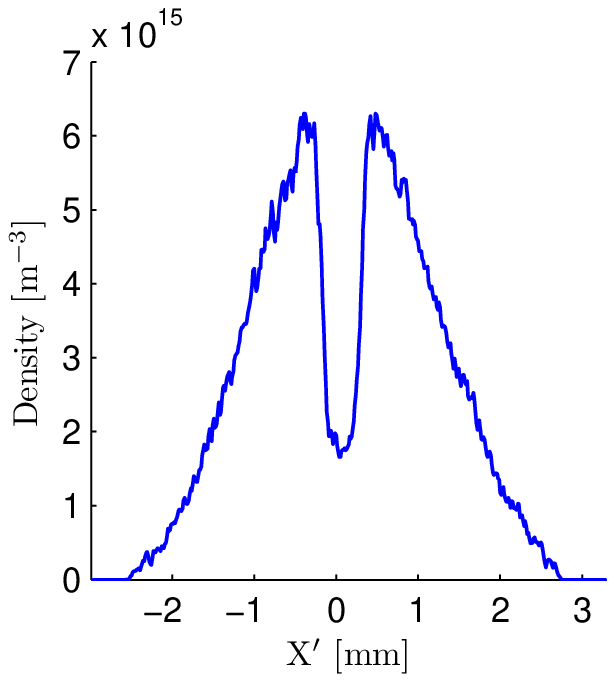}
\label{fig:streamplasma}}
\caption{\textbf{Streaming plasma}
(left)
2D image of two plasma hemispheres. An opaque wire is used as a transmission mask, producing a gap in the middle of the plasmas.
(right)
1D density profile of two plasma hemisphere. Ions are accelerated strongly into the void were they can interact and penetrate into the opposite hemisphere.}
\label{fig:stream}
\end{figure}
\end{center}

A simple configuration is the total attenuation of half the ionizing beam. This creates a single, hard-edged, hemisphere with rapid expansion perpendicular to the edge.  Figure \ref{fig:hemi} shows this configuration and a clear picture of the maximum density gradient currently observable. Further studies will allow us to characterize  this effect and possibly observe strongly supersonic velocities that may produce shock waves.

Another interesting configuration utilizes a single opaque wire for the transmission mask, thus creating two plasma hemispheres with an adjustable gap in between (See Fig. \ref{fig:stream}). In this geometry, ions accelerate into the central void, where they can interact. Preliminary studies have shown a strong dependance on experimental parameters (density, gradient and electron temperature)  of the ability for streaming plasmas to penetrate each other. Additional studies will  probe  subsonic and supersonic plasma velocities and look for effects of strong coupling.

\section{Conclusion}
A new relay lens system on the photoionization laser beam has resulted in at least a five-fold improvement in ionizing resolution compared to previous studies of sculpted UCNPs \cite{cmk10}, allowing access to many interesting plasma experiments.
We have measured shorter wavelength IAWs with very fast damping and observed ions that are accelerated strongly by a sharp plasma gradient.
Resolution of the imaging system may currently limit the observable resolution, suggesting the underlying density features are significantly sharper. 


\section{References}
\bibliography{bibliography}

\begin{thebibliography}{10}

\bibitem{sti92}
T.~H. Stix.
\newblock {\em Waves in Plasmas}.
\newblock AIP, New York, 2nd edition, 1992.

\bibitem{ich04}
S.~Ichimuru.
\newblock {\em Statistical {P}lasma {P}hysics, Volume II: Condensed Plasmas},
  volume~2 of {\em Frontiers in Physics}.
\newblock Westview Press, Boulder, CO, 2004.

\bibitem{mel86}
D.~B. Melrose.
\newblock {\em Instabilities in Space and Laboratory Plasmas}.
\newblock Cambridge University Press, Cambridge, 1986.

\bibitem{ich82}
S.~Ichimuru.
\newblock Strongly coupled plasmas: high-density classical plasmas and
  degenerate electron liquids.
\newblock {\em { Rev. Mod. Phys.}}, 54(4):1017, 1982.

\bibitem{mur04}
M.~S. Murillo.
\newblock Strongly coupled plasma physics and high energy density matter.
\newblock {\em Phys.\ Plasmas}, 11:2964, 2004.

\bibitem{cmk10}
J.~Castro, P.~McQuillen, and T.~C. Killian.
\newblock Ion acoustic waves in ultracold neutral plasmas.
\newblock {\em {Phys. Rev. Lett}}, 105:065004, 2010.

\bibitem{kpp07}
T.~C. Killian, T.~Pattard, T.~Pohl, and J.~M. Rost.
\newblock Ultracold neutral plasmas.
\newblock {\em { Phys. Rep.}}, 449:77, 2007.

\bibitem{kil07}
T.~C. Killian.
\newblock Ultracold neutral plasmas.
\newblock {\em {Science}}, 316:705, 2007.

\bibitem{mur01}
M.~S. Murillo.
\newblock Using {F}ermi statistics to create strongly coupled ion plasmas in
  atom traps.
\newblock {\em {Phys. Rev. Lett.}}, 87(11):115003, 2001.

\bibitem{scg04}
C.~E. Simien, Y.~C. Chen, P.~Gupta, S.~Laha, Y.~N. Martinez, P.~G. Mickelson,
  S.~B. Nagel, and T.~C. Killian.
\newblock Using absorption imaging to study ion dynamics in an ultracold
  neutral plasma.
\newblock {\em {Phys. Rev. Lett.}}, 92(14):143001, 2004.

\bibitem{csl04}
Y.~C. Chen, C.~E. Simien, S.~Laha, P.~Gupta, Y.~N. Martinez, P.~G. Mickelson,
  S.~B. Nagel, and T.~C. Killian.
\newblock Electron screening and kinetic energy oscillations in a strongly
  coupled plasma.
\newblock {\em { Phys. Rev. Lett.}}, 93:265003, 2004.

\bibitem{cdd05}
E.~A. Cummings, J.~E. Daily, D.~S. Durfee, and S.~D. Bergeson.
\newblock Fluorescence measurements of expanding strongly-coupled neutral
  plasmas.
\newblock {\em { Phys. Rev. Lett.}}, 95:235001, 2005.

\bibitem{mur06PRL}
M.~S. Murillo.
\newblock Ultrafast dynamics of strongly coupled plasmas.
\newblock {\em {Phys. Rev. Lett.}}, 96:165001, 2006.

\bibitem{ppr05PRL}
T.~Pohl, T.~Pattard, and J.~M. Rost.
\newblock Relaxation to non-equilibrium in expanding ultrcold neutral plasmas.
\newblock {\em { Phys. Rev. Lett.}}, 94:205003, 2005.

\bibitem{rha03}
F.~Robicheaux and J.~D. Hanson.
\newblock Simulated expansion of an ultra-cold, neutral plasma.
\newblock {\em { Phys. Plasmas}}, 10(6):2217, 2003.

\bibitem{kon02}
S.~G. Kuzmin and T.~M. O'Neil.
\newblock Numerical simulation of ultracold plasmas.
\newblock {\em { Phys. Plasmas}}, 9(9):3743, 2002.

\bibitem{mck02}
S.~Mazevet, L.~A. Collins, and J.~D. Kress.
\newblock Evolution of ultracold neutral plasmas.
\newblock {\em { Phys. Rev. Lett.}}, 88(5):55001, 2002.

\bibitem{gls07}
P.~Gupta, S.~Laha, C.~E. Simien, H.~Gao, J.~Castro, T.~C. Killian, and T.~Pohl.
\newblock Electron temperature evolution in expanding ultracold neutral
  plasmas.
\newblock {\em { Phys. Rev. Lett.}}, 99:75005, 2007.

\bibitem{fzr07}
R.~S. Fletcher, X.~L. Zhang, and S.~L. Rolston.
\newblock Using three-body recombination to extract electron temperatures of
  ultracold plasmas.
\newblock {\em Phys. Rev. Lett.}, 99(14):145001, 2007.

\bibitem{zfr08expansion}
X.~L. Zhang, R.~S. Fletcher, S.~L. Rolston, P.~N. Guzdar, and M.~Swisdak.
\newblock Ultracold plasma expansion in a magnetic field.
\newblock {\em Phys. Rev. Lett.}, 100(23):235002, 2008.

\bibitem{lgs07}
S.~Laha, P.~Gupta, C.~E. Simien, H.~Gao, J.~Castro, and T.~C. Killian.
\newblock Experimental realization of an exact solution to the {V}lasov
  equations for an expanding plasma.
\newblock {\em {Phys. Rev. Lett.}}, 99:155001, 2007.

\bibitem{rha02}
F.~Robicheaux and J.~D. Hanson.
\newblock Simulation of the expansion of an ultracold neutral plasma.
\newblock {\em { Phys. Rev. Lett.}}, 88(5):55002, 2002.

\bibitem{klk01}
T.~C. Killian, M.~J. Lim, S.~Kulin, R.~Dumke, S.~D. Bergeson, and S.~L.
  Rolston.
\newblock Formation of {R}ydberg atoms in an expanding ultracold neutral
  plasma.
\newblock {\em { Phys. Rev. Lett.}}, 86(17):3759, 2001.

\bibitem{pvs08}
T.~Pohl, D.~Vrinceanu, and H.~R. Sadeghpour.
\newblock Rydberg atom formation in ultracold plasmas: Small energy transfer
  with large consequences.
\newblock {\em Phys. Rev. Lett.}, 100:223201, 2008.

\bibitem{gpr03}
T.~F. Gallagher, P.~Pillet, M.~P. Robinson, B.~Laburthe-Tolra, and M.~W. Noel.
\newblock Back and forth between {R}ydberg atoms and ultracold plasmas.
\newblock {\em { J. Opt. Soc. Am B}}, 20(5):1091, 2003.

\bibitem{ccb05}
F.~Cornolti, F.~Ceccherini, S.~Betti, and F.~Pegoraro.
\newblock Charged state of a spherical plasma in vacuum.
\newblock {\em Phys. Rev. E}, 71(5):056407, 2005.

\bibitem{lcg06}
S.~Laha, Y.~C. Chen, P.~Gupta, C.~E. Simien, Y.~N. Martinez, P.~G. Mickelson,
  S.~B. Nagel, and T.~C. Killian.
\newblock Kinetic energy oscillations in annular regions of ultracold neutral
  plasmas.
\newblock {\em {Euro. Phys. J. D}}, 40:51, 2006.

\bibitem{kkb00}
S.~Kulin, T.~C. Killian, S.~D. Bergeson, and S.~L. Rolston.
\newblock Plasma oscillations and expansion of an ultracold neutral plasma.
\newblock {\em { Phys. Rev. Lett.}}, 85(2):318, 2000.

\bibitem{fzr06}
R.~S. Fletcher, X.~L. Zhang, and S.~L. Rolston.
\newblock Observation of collective modes of ultracold plasmas.
\newblock {\em { Phys. Rev. Lett.}}, 96:105003, 2006.

\bibitem{zfr08}
X.~L. Zhang, R.~S. Fletcher, and S.~L. Rolston.
\newblock Observation of an ultracold plasma instability.
\newblock {\em Phys. Rev. Lett.}, 101(19):195002, 2008.

\bibitem{wbh97}
X.~Wang and A.~Bhattacharjee.
\newblock Hydrodynamic waves and correlation functions in dusty plasmas.
\newblock {\em Phys. Plasmas}, 4(11):3759, 1997.

\bibitem{rka97}
M.~Rosenberg and G.~Kalman.
\newblock Dust acoustic waves in strongly coupled dusty plasmas.
\newblock {\em Phys. Rev. E}, 56(6):7166, 1997.

\bibitem{kse98}
P.~K. Kaw and A.~Sen.
\newblock Low frequency modes in strongly coupled dusty plasmas.
\newblock {\em Phys. Plasmas}, 5(10):3552, 1998.

\bibitem{mur98}
M.~S. Murillo.
\newblock Static local field correction description of acoustic waves in
  strongly coupling dusty plasmas.
\newblock {\em Phys. Plasmas}, 5(9):3116, 1998.

\bibitem{mur00PPlasma}
M.~S. Murillo.
\newblock Longitudinal collective modes of strongly coupled dusty plasmas at
  finite frequencies and wavevectors.
\newblock {\em Phys. Plasmas}, 7(1):33, 2000.

\bibitem{kaw01}
P.~K. Kaw.
\newblock Collective modes in a strongly coupled dusty plasma.
\newblock {\em Phys. Plasmas}, 8(5):1870, 2001.

\bibitem{ber93}
M.~A. Berkovsky.
\newblock Ion modes in strongly coupled two-component plasmas.
\newblock {\em J. Plasma Phys.}, 50:359, 1993.

\bibitem{nbs99}
Y.~Nakamura, H.~Bailung, and P.~K. Shukla.
\newblock Observation of ion-acoustic shocks in a dusty plasma.
\newblock {\em Phys. Rev. Lett.}, 83(8):1602, 1999.

\bibitem{cgk08}
J.~Castro, H.~Gao, and T.~C. Killian.
\newblock Using sheet fluorescence to probe ion dynamics in an ultracold
  neutral plasma.
\newblock {\em Plasma Phys.\ Control.\ Fusion}, 50:124011, 2008.

\bibitem{bth10}
R.~Blandford and K.~Thorne.
\newblock {\em Application of Classical Physics}.
\newblock in preparation,
  http://www.pma.caltech.edu/Courses/ph136/yr2008/0820.1.K.pdf, 2010.

\end{thebibliography}
\include{bibliography}

\end{document}